\documentclass[psfig]{aastex}

\usepackage{emulateapj5}

\def\ltsima{$\; \buildrel < \over \sim \;$}

\def\lsim{\lower.5ex\hbox{\ltsima}}
\def\gtsima{$\; \buildrel > \over \sim \;$}
\def\gsim{\lower.5ex\hbox{\gtsima}}

\begin{document}
\title{Redshifted 21cm Signatures Around the Highest Redshift Quasars}

\author{J. Stuart B. Wyithe\altaffilmark{1} and Abraham
Loeb\altaffilmark{2}}

\email{swyithe@isis.ph.unimelb.edu.au; aloeb@cfa.harvard.edu}

\altaffiltext{1}{University of Melbourne, Parkville, Victoria, Australia}

\altaffiltext{2}{Harvard-Smithsonian Center for Astrophysics, 60 Garden
St., Cambridge, MA 02138}

\begin{abstract}
The Ly$\alpha$ absorption spectrum of the highest redshift quasars
indicates that they are surrounded by giant \ion{H}{2} regions, a few Mpc
in size.  The neutral gas around these \ion{H}{2} regions should emit 21cm
radiation in excess of the Cosmic Microwave Background, and enable future
radio telescopes to measure the transverse extent of these \ion{H}{2}
regions.  At early times, the \ion{H}{2} regions expand with a relativistic
speed. Consequently, their measured sizes along the line-of-sight (via
Ly$\alpha$ absorption) and transverse to it (via 21 cm emission) should
have different observed values due to relativistic time-delay.  We show
that the combined measurement of these sizes would directly constrain the
neutral fraction of the surrounding intergalactic medium (IGM) as well as
the quasar lifetime.  Based on current number counts of luminous quasars at
$z\ga6$, an instrument like {\em LOFAR} should detect $\ga2$ redshifted
21cm shells per field (with a radius of $11^\circ$) around active quasars
as bright as those already discovered by SDSS, and $\ga200$ relic shells of
inactive quasars per field. We show that Ly$\alpha$ photons from the quasar
are unable to heat the IGM or to couple the spin and kinetic temperatures
of atomic hydrogen beyond the edge of the \ion{H}{2} region. The detection
of the IGM in 21cm emission around high redshift quasars would therefore
gauge the presence of a cosmic Ly$\alpha$ background during the
reionization epoch.

\end{abstract}


\section{Introduction}

The recent discovery of quasars at $z>6$ (Fan et al.~2001; Fan et al.~2003)
and the subsequent observation of two Gunn-Peterson~(1965) troughs
(Djorgovski et al.~2003; White et al.~2003) established a new era in
studies of the end of the reionization epoch.  The absorption of Ly$\alpha$
photons from these quasars probes the ionization state of the hydrogen in
the intergalactic medium (IGM) near the quasar redshift and has revealed
giant ionized regions (Cen \& Haiman~2000; Madau \& Rees~2000). Comparison
of the radii of these regions with the ionizing flux implies quasar ages of
$\la10^7$ years, and an ionization front that is expanding with a
relativistic speed (Wyithe \& Loeb 2004).

The generation of planned low frequency instruments (such as {\it
LOFAR}\footnote{See http://www.lofar.org} or the {\it SKA}\footnote{ See
http://www.skatelescope.org}) will open a new window on the state of
hydrogen in the pre-reionization universe through the emission or
absorption of 21cm photons relative to the Cosmic Microwave Background
(CMB) by neutral hydrogen in the IGM (Scott \& Rees 1990; Madau, Meiksin \&
Rees~1997; Gnedin \& Ostriker 1997; Shaver et al. 1999; Tozzi et al.~2000;
Iliev et al. 2002, 2003; Ciardi \& Madau 2003; Furlanetto, Sokasian \&
Hernquist 2003; Zaldarriaga, Furlanetto, \& Hernquist 2003; Gnedin \&
Shaver 2003; Loeb \& Zaldarriaga 2003).  The giant \ion{H}{2} regions surrounding
the highest redshift quasars should be visible via rings of redshifted 21cm
emission from warm but neutral gas beyond the ionization front.

The relativistic expansion of the \ion{H}{2} region implies that the radii
measured along and transverse to the line-of-sight using the Ly$\alpha$ and
21cm techniques respectively, probe different epochs due to relativistic
time delay.  In this paper we discuss what will be learned about the state
of the IGM surrounding the highest redshift quasars by combining Ly$\alpha$ and
21cm measurements. In \S~\ref{SSp} we discuss the measurements of radii of
the \ion{H}{2} region in Ly$\alpha$ absorption and redshifted 21cm emission. The
relation between these radii and the neutral fraction of the IGM are then
discussed in \S~\ref{comb}. Some examples of the brightness temperature
profiles around high redshift quasars are presented in \S~\ref{ex}. In
\S~\ref{fossil} we discuss the observation of fossil \ion{H}{2} regions and their
utility as a probe of the radiation background at high redshift. Finally in
\S~\ref{global} we discuss the utility of quasar Str\"omgren spheres in the
calibration of the global signature of reionization.  \S~\ref{disc}
summarizes our main results. Throughout the paper we adopt the set of
cosmological parameters determined by the {\em Wilkinson Microwave Anisotropy Probe}
(WMAP, Spergel et al. 2003), namely mass density
parameters of $\Omega_{m}=0.27$ in matter, $\Omega_{b}=0.044$ in baryons,
$\Omega_\Lambda=0.73$ in a cosmological constant, and a Hubble constant of
$H_0=71~{\rm km\,s^{-1}\,Mpc^{-1}}$.

\section{Observations of Str\"omgren Spheres in Redshifted 21cm 
Emission and Ly$\alpha$ Absorption}
\label{SSp}

First, we consider the observed signatures of \ion{H}{2} regions (Str\"omgren
spheres) around the highest redshift quasars during their early
relativistic expansion.

\subsection{Evolution of the Str\"omgren Sphere Around a High Redshift Quasar}

Neglecting recombinations which do not become important until the expansion
becomes sub-relativistic, White et al.~(2003) derived the evolution of the
physical radius of the Str\"omgren Sphere ($R_{\rm p}$) as a function of
the source age ($t_{\rm age}$) which can be easily extended to include the
possibility of a non-zero radius ($R_{\rm 0}$) at the initial time $t_{\rm
age}=0$,
\begin{equation}
\label{SE1}
\dot{{N}}\left(t_{\rm age}-\frac{R_{\rm p}}{c}\right) = \frac{4\pi
}{3}\left(R_{\rm p}{^3}-R_{\rm 0}{^3}\right)x_{\rm HI}n_{\rm H}^0(1+z)^3,
\end{equation}
where $\dot{{N}}$ is the production rate of ionizing photons by
the quasar, $n_{\rm H}^0$ is the co-moving hydrogen number density, $x_{\rm
HI}$ is the neutral fraction and $c$ is the speed of light.  This cubic
equation may be solved algebraically (White et al.~2003).  Alternatively,
$R_p$ may be found by integrating the differential equation
\begin{equation}
\label{SE2}
\frac{dR_{\rm p}}{dt}=c\left(\frac{\dot{{N}}}{\dot{{N}}+4\pi R_{\rm
p}^2cx_{\rm HI}n_{\rm H}^0(1+z)^3}\right),
\end{equation}
with the initial condition $R_{\rm p}(0)=R_0$.  The latter approach (Wyithe
\& Loeb~2004) allows for the inclusion of recombinations. As long as the
number of photons emitted per time interval $\Delta t$ is larger than the
number of hydrogens in a shell of thickness $\Delta R = c \Delta t$ at
$R_{\rm p}$, the \ion{H}{1} region continues to expand at nearly the speed of
light.

Calculations of $R_{\rm p}$ in this paper assume that the ionization front
is thin. The spectrum averaged mean-free path for the ionizing quasar
photons is $\lambda\sim 1.5x_{\rm HI}^{-1}(1+z)^{-3}$Mpc, which may be
compared to the bubble radius $R_{\rm p}\sim4.5$Mpc, to yield the
fractional thickness of the ionization front $f\equiv ({\lambda}/{R_{\rm
p}})\sim 2\times10^{-3}\left({R_{\rm p}}/{4.5{\rm Mpc}}\right)^{-1}x_{\rm
HI}^{-1}\left[{(1+z)}/{7.3}\right]^{-3}$.  The assumption of a thin
ionization front may be considered reasonable provided that $x_{\rm
HI}\gsim 10^{-2}$. A low neutral fraction is unlikely given the size of the
observed \ion{H}{2} regions ($R_{\rm p}\sim4.5$Mpc) around the $z\ga 6.3$
quasars (Wyithe \& Loeb 2004). X-ray photons have a longer mean free path
but are rare and do not contribute significantly to the number of
ionizations.

\subsection{The Str\"omgren Sphere Radius 
Measured Through Ly$\alpha$ Absorption}

Equations~(\ref{SE1}) and (\ref{SE2}) refer to the Str\"omgren sphere
radius at a time $t_{\rm age}$ after the source turns on. However
measurements of the radius through Ly$\alpha$ absorption ($R_{\rm Ly}$) are
made along the line of sight.  Ly$\alpha$ photons emitted at $t_{\rm age}$
are not absorbed until a time $t$
satisfying the relation $t=t_{\rm age}+R(t)/c$ (note that the quasar may no
longer be active at $t$). When considering the value of $R_{\rm Ly}$
measured through Ly$\alpha$ absorption, the relevant time to consider is
therefore the age of the quasar at the time when the photons being absorbed
at the edge of the \ion{H}{2} region were emitted $t_{\rm age,e}=t_{\rm
age}-R(t_{\rm age})/c$. In the absence of recombinations (White et
al.~2003), the relation between $R_{\rm Ly}$ and $t_{\rm age,e}$ is
equivalent to the infinite speed of light solution for the evolution of a
Str\"omgren sphere radius (Haiman \& Cen~2000; Madau \& Rees~2000),
\begin{eqnarray}
\label{evinf}
\nonumber R_{\rm
Ly}&=&\left(R_0^3+\frac{3}{4\pi}\frac{\dot{N}_\parallel t_{\rm
age,e}}{x_{\rm HI}n_{\rm
H}^0(1+z)^3}\right)^{1/3}=4\mbox{Mpc}\left[\left(\frac{R_0}{3.2\mbox{Mpc}}\right)^3\right.\\
\nonumber &&\hspace{10mm}+\left.
\left(\frac{\dot{N}_\parallel }{10^{57}\mbox{s}^{-1}}\right)\left(\frac{t_{\rm
age,e}}{10^7\mbox{yr}}\right)\left(\frac{1+z}{7.5}\right)^{-3}\right]^{1/3}
\end{eqnarray}
where we add the subscript $\parallel$ to denote the isotropic equivalent
value of $\dot{N}$ for a measurement done parallel to the line-of-sight.

\subsection{The Str\"omgren radius measured through \ion{H}{1} emission and absorption}
\label{SSEA}

Madau, Meiksin \& Rees~(1997), and Tozzi et al.~(2000) have considered the
\ion{H}{1} signature of the IGM surrounding a high redshift quasar. Outside the
\ion{H}{2} region, neutral hydrogen may be heated by soft X-rays and by
scattering of Ly$\alpha$ photons.  The spin temperature of the neutral
hydrogen is coupled to the kinetic temperature of the IGM through
Ly$\alpha$ scattering of a background of UV photons, 
and results in 21cm emission (in excess of the CMB flux) around the quasar.

\subsubsection{The Ly$\alpha$ Scattering rate}

The spin temperature of a warm IGM quickly becomes coupled to the CMB
radiation, so that it yields no contrast for \ion{H}{1} observations (Madau, Meiksin
\& Rees~1997).  However scatterings of Ly$\alpha$ photons couple the spin
temperature of the warm IGM at radii $R>R_{\rm p}$ through the
Wouthuysen-Field effect (Wouthuysen~1952; Field~1958). The spin temperature
\begin{equation}
T_{\rm S}=\frac{T_{\rm CMB}+y_\alpha T_{\rm IGM}}{1+y_\alpha}
\end{equation}
is weighted between the CMB temperature and
the IGM temperature through the Ly$\alpha$ pumping efficiency 
\begin{equation}
\label{couple}
y_\alpha  \equiv \frac{4P_\alpha T_\star}{27A_{10}T_{\rm IGM}}.
\end{equation}
In equation~(\ref{couple}), $T_\star=0.068$K is the temperature
corresponding to the transition energy, $A_{10}=2.85\times10^{-15}$s$^{-1}$
is the spontaneous decay rate of the hyperfine transition, and $P_\alpha$
is the Ly$\alpha$ scattering rate per hydrogen.

The value of $P_\alpha$ may be calculated for a continuum UV source
embedded in an expanding IGM around a Str\"omgren sphere as follows (see
Loeb \& Rybicki 1999 for more details on the related physics).  In the
absence of scattering, photons that start at a frequency separation $\Delta
\nu$ on the blue side of the Ly$\alpha$ resonance frequency $\nu_\alpha$
would redshift into resonance at some distance $R_{\rm res}=(\Delta
\nu/\nu_\alpha)c/H$ from the source, where $H=H(z)$ is the Hubble parameter
at the quasar redshift. At this distance, the Hubble velocity of the IGM
introduces a Doppler shift that is equal to the initial blueshift of these
photons relative to the Ly$\alpha$ line center. When a resonant photon
enters the neutral IGM near $R_{\rm res}$, its mean-free-path becomes very
short because of the large cross-section to Ly$\alpha$ scattering. As the
photon frequency redshifts very close to the Ly$\alpha$ resonance, the
mean-free-path diminishes, and the frequency shift per scattering (which
corresponds to the Hubble velocity increment across a mean-free-path) goes
to zero. Consequently, the photon spends a very long time around $R_{\rm
res}$. At this stage, the finite temperature of the gas becomes
important. The photons are able to diffuse out of resonance primarily due
to the Doppler shifts they encounter by the thermal speed of the atoms at
$R_{\rm res}$.

Each scattering near $R_{\rm res}$ leads to a fractional frequency shift by
the thermal Doppler effect of $\Delta \nu_{\rm th}/\nu_\alpha=v_{\rm
th}/c$, where $v_{\rm th}=(2kT_{\rm IGM}/m_p)^{1/2}$ is the thermal speed
of the atoms and $\nu_\alpha$ is the resonant Ly$\alpha$ frequency. The
thermal shifts are of random sign and so the photons go through a random
walk in frequency. Consider the photons just around the resonant frequency
at $R_{\rm res}$. These photons change their frequency by $\sim\Delta
\nu_{\rm th}$ in each scattering, and have a mean free path of
$\lambda_{\rm th}=1/[n_{\rm H}^0(1+z)^3\sigma_{\rm th}]$. 
Here $\sigma_{\rm th}$ is the Ly$\alpha$ cross-section at the center of the
thermally broadened Ly$\alpha$ line (Rybicki \& Lightman~1979)
\begin{equation}
\sigma_{\rm th} = \frac{\sqrt{\pi} e^2}{m_{\rm e}c}\frac{f_{12}}{\Delta
\nu_{\rm th}}=5.87\times10^{-12}\mbox{cm}^2\left(T_{\rm
IGM}\over1~\mbox{K}\right)^{-1/2},
\end{equation}
where $e$ and $m_{\rm e}$ are the electron charge and mass, and $f_{21}$ is
the oscillator strength for the Ly$\alpha$ transition.  The mean intensity
of nearly resonant photons at $R_{\rm res}$ is increased by a factor of
$\sim l/\lambda_{\rm th}$ relative to optically-thin conditions, where $l$
is the so-called Sobolev length, defined as the distance over which the
bulk velocity gradient of the medium equals the thermal speed of an atom,
\begin{equation}
l=\frac{v_{\rm th}}{dv/dR}=\frac{v_{\rm th}}{H},
\end{equation}
and $v=HR$ is the Hubble velocity.  The optical-depth factor $\tau_{\rm
th}=l/\lambda_{\rm th}$ reflects the fact that the diffusion speed of
photons across the resonance region of width $\sim l$ is a factor
$\lambda_{\rm th}/l$ times smaller than the speed of light.

In a steady state, the mean intensity at all radii should therefore be
enhanced relative to the optically-thin case by a factor of $\sim \tau_{\rm
th}$.  However, a steady state will not be established immediately when
light from the quasar first penetrates the neutral IGM because of the slow
diffusion speed of the resonant photons.  At early times, scattering will
produce an exponential reduction in resonant photon flux by a factor
$\propto e^{-\tau_{\rm th}}$.  Even though photons may be sufficiently blue
that they do not redshift into the resonance until $R_{\rm res}$, they
would still be absorbed by the blue wing of the Ly$\alpha$ resonance (where
the cross-section is still considerable) at radii smaller than
$R_{\rm res}$. 

Let us therefore consider photons of arbitrary frequency on
the blue side of the Ly$\alpha$ resonance and calculate their mean
intensity at $R_{\rm res}$. We define the mean intensity as the angular
average of the specific (directional) intensity $I(\nu,\mu)$, namely
$J(\nu)={1\over 2}\int_{-1}^{1}d\mu I(\nu,\mu)$ where $\mu=\cos\theta$ is
the direction relative to the radius vector at $R_{\rm res}$ (Loeb \&
Rybicki 1999).
The Ly$\alpha$ optical depth for photons with frequency $\nu$ at a radius
$R_{\rm res}$ is
\begin{equation}
\tau(\nu)=\int_{R_{\rm res}-\Delta R_{\rm pl}}^{R_{\rm res}} dR'
\sigma_{\alpha}\left(\nu[1+\frac{H}{c}(R_{\rm res}-R')]\right)n^0_{\rm
H}(1+z)^3,
\end{equation}
where $\sigma_\alpha(\nu)=\sigma_{\rm
th}\exp\{-(\nu-\nu_\alpha)^2/\Delta\nu_{\rm th}^2\}$ is the
thermally-broadened cross-section (Rybicki \& Lightman 1979), 
and $\Delta R_{\rm pl}=R_{\rm res}-R_{\rm p}(t-\Delta t)$ is the
path-length traversed by a photon through the neutral IGM after it had
crossed the boundary of the \ion{H}{2} region at a time $t-\Delta t$.  In
the strong scattering regime ($\tau(\nu)\gg 1$), the mean intensity is a
factor of $\tau(\nu)$ larger than the specific intensity along the radial
direction, because the radial diffusion speed of the photons is $\sim
c/\tau(\nu)$.  The initial mean intensity of nearly resonant photons at
radius $R_{\rm res}$ is therefore
\begin{equation}
\label{I}
J(\nu)\sim \frac{1}{4\pi R_{\rm res}^2}\epsilon_{\rm
local}(\nu)\,\tau(\nu) e^{-\tau(\nu)},
\end{equation}
where 
\begin{equation}
\epsilon_{\rm local}(\nu) = \epsilon\left(\nu(1+\frac{HR_{\rm
res}}{c})\right)\left[\frac{HR_{\rm res}}{c}+1\right]^{-1}
\end{equation}
is the redshifted source spectrum $\epsilon(\nu)$ in ${\rm
erg~s^{-1}~Hz^{-1}}$.  The scattering rate $P_\alpha$ may then be
calculated from the mean intensity as
\begin{equation}
P_\alpha = \int_0^\infty d\nu \frac{J(\nu)}{h_{\rm p}\nu} \sigma_\alpha (\nu).
\end{equation}

The directional intensity of resonant Ly$\alpha$ photons starts high near
the boundary of the \ion{H}{2} region at $R_{\rm p}$, but is initially
suppressed at larger radii.  Resonant photons will diffuse to radii larger
than $R_{\rm p}$ at a speed $\sim c/\tau_{\rm th}$. In the case of a
stationary \ion{H}{2} front, the high intensity region expands with time,
extending a distance $(c/\tau_{\rm th})\Delta t$ beyond the edge of the
\ion{H}{2} region after a time $\Delta t$, and absorption will truncate the
intensity at radii larger than $R_{\rm p}+(c/\tau_{\rm th})\Delta t$.  For
the quasars under consideration the expansion speed of the \ion{H}{2}
region ($\ga 0.1 c$) is well in excess of the diffusion speed\footnote{Note
that the diffusion speed of a photon is a function of time as its frequency
drifts in and out of resonance with the Ly$\alpha$ transition of the
expanding IGM.} of resonant Ly$\alpha$ photons ($c/\tau_{\rm
th}\sim10^{-5}c$).  On the timescales of interest here, there
is a deficit of photons near the line center as described by
equation~(\ref{I}).

The critical thermalization rate above which Ly$\alpha$ scattering 
couples $T_{\rm S}$ to $T_{\rm IGM}$ is given by (Madau, Meiksin \&
Rees~1997)
\begin{equation}
P_{\rm th}\equiv \frac{27A_{10}T_{\rm
CMB}}{4T_\star}\sim5\times10^{-12}\left(\frac{1+z}{7.5}\right)\mbox{s}^{-1}
\end{equation}
We find that $P_\alpha$ only exceeds the thermalization rate $P_{\rm th}$
within a negligibly thin shell around the \ion{H}{2} region
($\la10^{-4}$Mpc).
As a result, detection of redshifted 21cm emission from the \ion{H}{2}
regions of quasars will require the presence of a UV background.
Background photons between the Ly$\alpha$ and the Lyman-limit frequencies
can propagate through the universe and redshift with time to the point
where they resonate with the Ly$\alpha$ line and couple $T_{\rm s}$ to
$T_{\rm K}$.  Such a background is expected to be present at $z\sim6.5$
since the large electron scattering optical depth measured by WMAP 
(Kogut et al.~2003) implies that the
IGM had been significantly reionized by that time (requiring the production
of more than one ionizing photon per baryon above the Lyman-limit).

Our result for the scattering rate should be compared with the result of
Madau, Meiksin \& Rees~(1997) who estimated
\begin{eqnarray}
P_\alpha &=& \int_0^\infty d\nu \frac{\epsilon(\nu)}{h_{\rm
p}\nu}\frac{1}{4\pi R^2}\sigma_\alpha(\nu) \\ \nonumber &\sim&
\left(8\times10^{-11}\mbox{s}^{-1}\right)\left(\frac{R}{1\mbox{Mpc}}\right)^{-2}\left(\frac{\nu_{\alpha}\epsilon(\nu_{\alpha})}{3\times10^{46}\mbox{erg}\,\mbox{s}^{-1}\,\mbox{Hz}^{-1}}\right).
\end{eqnarray}
In the steady state this expression incorrectly neglects the rise in
intensity that follows from the small diffusion speed of photons near the
Ly$\alpha$ resonance and hence underestimates the value of $P_\alpha$ by a
factor of $\sim n_{\rm H}^0(1+z)^3\sigma_{\rm th}v_{\rm th}/H$ which
amounts to several orders of magnitude. However, in the initial state the
expression neglects the absorption of resonant photons before they reach a
radius $R>R_{\rm p}$, and so overestimates the value of $P_\alpha$ outside the
\ion{H}{2} region by several orders of magnitude.

The scattering of Ly$\alpha$ photons may also provide a source of heating
for the gas.  As this peak shifts to longer wavelengths, 
it loses energy to the hydrogen atoms off which the photons scatter
(Chen \& Miralda-Escude~2003). The additional energy density due to
the scattering is
\begin{equation}
U = \frac{\epsilon_{\rm local}(\nu_\alpha)}{4\pi R^2
c}\left(\frac{l}{\lambda}-1\right)\Delta \nu_{\rm th} \sim
P_\alpha\frac{h_{\rm p}\nu_\alpha}{c\sigma_{\rm th}}.
\end{equation}
The rate at which energy is lost from the line in an expanding IGM equals 
the heating rate of the
gas. Hence we have
\begin{equation}
\Gamma_\alpha = \frac{dU}{dt} =
U \frac{1}{\nu}\frac{d\nu}{dt}=HP_\alpha\frac{h_{\rm
p}\nu_\alpha}{c\sigma_{\rm th}}.
\end{equation}
This heating rate, which is proportional to the Ly$\alpha$ scattering rate,
does not contribute to the heating of gas outside the \ion{H}{2} region owing to
the very low Ly$\alpha$ intensity there.  Instead, the heating is dominated
by X-rays which we discuss next.

\subsubsection{X-ray Heating Beyond the \ion{H}{2} Region}

Gas in the IGM outside the ionized region is subject to secondary electron
heating resulting from ionizations by soft X-rays that propagate into the
neutral IGM beyond the \ion{H}{2} region.
The heating rate $\Gamma_{\rm x}$ in $\rm erg~s^{-1}~Mpc^{-3}$ at radius
$R>R_{\rm p}$ due to a source with a spectrum $\epsilon(\nu)$ (in
erg/sec/Hz) is
\begin{eqnarray}
\nonumber
\label{heat}
\Gamma_{\rm X}(R,t) &=& \int_{\nu_{\rm ion}}^\infty d\nu f_{\rm
x}\frac{\epsilon_{\rm local}(\nu)}{4\pi R^2}\sigma_{\rm pi}(\nu)x_{\rm
HI}n_{H}^0(1+z)^3\\ &\times&\exp\left[-\Delta R_{\rm pl}\sigma_{\rm
pi}(\nu)x_{\rm HI}n_{H}^0(1+z)^3\right],
\end{eqnarray}
where $\sigma_{\rm pi}(\nu)$ is the cross-section for photoionization and
$\nu_{\rm ion}=3.29\times10^{15}\mbox{Hz}$ is the Lyman-limit frequency
corresponding to the ionization threshold of hydrogen. In
equation~(\ref{heat}) $\Delta R_{\rm pl}=R-R_{\rm p}(t-\Delta t)$ is the
path-length of a photon through the neutral IGM after it crossed the
boundary of the \ion{H}{2} region at time $t-\Delta t$. We have neglected
adiabatic cooling due to cosmological expansion in this expression since
the quasar lifetime is much shorter than the Hubble time. The fraction of
photon energy converted into heat, $f_{\rm x}$,  may be
evaluated using the fitting formula (Shull \& van Steenberg~1985)
\begin{equation}
\label{fx}
f_{\rm x} = 0.9971\left(1-\left[1-(1-x_{\rm
HI})^{0.2663}\right]^{1.3163}\right).
\end{equation}
This equation is valid for photon energies $h_{\rm p}\nu \ga 100$eV. At
lower energies, the value of $f_{\rm x}$ may be roughly approximated by
multiplying equation~(\ref{fx}) with an additional factor of
$\exp{\left(\left[100\mbox{eV}-h_{\rm p}\nu \right]/100\mbox{eV}\right)}$
(with the constraint that $f_{\rm x}<1$).  The residual ionization fraction
from cosmological recombination of $\sim5\times10^{-4}$ yields $f_{\rm
x}\sim0.17$ (Madau et al.~1997), which sets the minimum heating rate.  The
heating rate leads to a temperature rise according to
\begin{equation}
\label{temp}
\frac{dT_{\rm IGM}}{dt} = \frac{2}{3}\frac{\Gamma_{\rm X}}{k_{\rm B}}\frac{1}{x_{\rm HI}n_{\rm H}^0(1+z)^3}.
\end{equation}

\subsubsection{Redshifted 21cm Observation of Quasar \ion{H}{2} Regions}

The above discussion has shown that X-rays from the quasar are capable of
heating the gas outside the \ion{H}{2} region, and that the spin
temperature of this warm gas can be coupled to its kinetic temperature in
the presence of a strong UV (Ly$\alpha$) background.  In regions of the IGM
where the spin temperature is decoupled from the CMB temperature, a
differential antenna temperature of
\begin{equation}
\label{delT}
\delta T_{\rm b}= (23\mbox{mK})x_{\rm
HI}\left(\frac{1+z}{7.5}\right)^{1/2}\left(\frac{T_{\rm S}-T_{\rm
CMB}}{T_{\rm S}}\right)
\end{equation}
is observed. In equation~(\ref{delT}) we have assumed a uniform IGM at the
mean density with a neutral fraction $x_{\rm HI}$. The luminous high
redshift quasars reside in rare high-density regions. However on the scale
of several Mpc, the overdensity due to infall around these regions is
$\la10\%$ (Barkana \& Loeb 2003a). 

The brightness temperature of the redshifted 21cm signal is small and
subject to severe foreground contamination problems. Studies by Di~Matteo
et al.~(2002), and Oh \& Mack~(2003) have found that on scales of several
arc-minutes, randomly distributed point sources would have to be removed
down to a level of $\sim$1$\mu$Jy before brightness temperature
fluctuations of order 10 mK could be detected. In the presence of the
expected level of clustering of these sources, even the removal of point
sources below 1$\mu$Jy would leave foreground fluctuations on this scale
that are several orders of magnitude in excess of the redshifted 21cm
signal. However, the spectrum of the foreground contaminants is expected to
be smooth (Shaver et al.~1999).  The width of the \ion{H}{2} region
corresponds to a recessional velocity $v\la 3000~{\rm km~s^{-1}}$ or a
redshift interval of $\sim(1+z)v/c\sim 0.07$.  The redshifted 21cm ring
will therefore produce a bump (or a dip) on top of the smooth foreground
spectrum at the redshift of the quasar.

At a frequency $\nu_{\rm bp}\sim 120$MHz, an instrument like {\it LOFAR} is
expected to have a band-pass of $\Delta \nu_{\rm bp}\sim4$MHz. This should
be compared with the frequency shift associated with the redshift
difference between light emitted at the quasar and the edge of the
\ion{H}{2} region along the line of sight
\begin{equation}
\Delta \nu \sim 2\mbox{MHz}\left(\frac{R_{\rm
Ly}}{4\mbox{Mpc}}\right)\left(\frac{1+z}{7.5}\right).
\end{equation}
Since many channels are contained within each band-pass, 
the structure of the \ion{H}{2} region can therefore be probed along the
line-of-sight in frequency space (Madau, Meiksin \& Rees~1997), allowing
determination of the full three dimensional shape of the \ion{H}{2} region,
including the radius of the \ion{H}{2} region behind the quasar in addition
to the radius $R_{\rm Ly}$ of the \ion{H}{2} region in front of the quasar.

There are two distinct possibilities for the redshifted 21cm signature of a
high redshift \ion{H}{2} region.  First, the region might be expanding into
a cold IGM, in which case emission will only be seen out to a radius where
the warming X-rays can reach. Alternatively, the IGM may have been
pre-heated, either by an X-ray background, by gravitationally induced
shocks in the IGM (Furlanetto \& Loeb~2003), or by an early reionization
(e.g. Wyithe \& Loeb~2003a; Cen~2003). The presence of Gunn-Peterson troughs
in quasar spectra at $z\ga6$ implies an IGM that is at least partially
neutral, while the relatively small sizes of the Str\"omgren spheres around
these quasars suggest a neutral fraction that is of order unity (Wyithe \&
Loeb~2004). On the other hand, observations from WMAP 
(Kogut et al.~2003) suggest an optical depth to
electron scattering of $\tau\sim0.17$ and therefore significant
reionization at $z\sim20$.  If the universe were reionized twice, then the
Str\"omgren spheres would be expanding into a partially neutral, but warm
IGM.

\section{comparison of Str\"omgren sphere radii measured using Ly$\alpha$ 
and \ion{H}{1} observations}
\label{comb}

\begin{figure*}[htbp]
\epsscale{2.}  \plotone{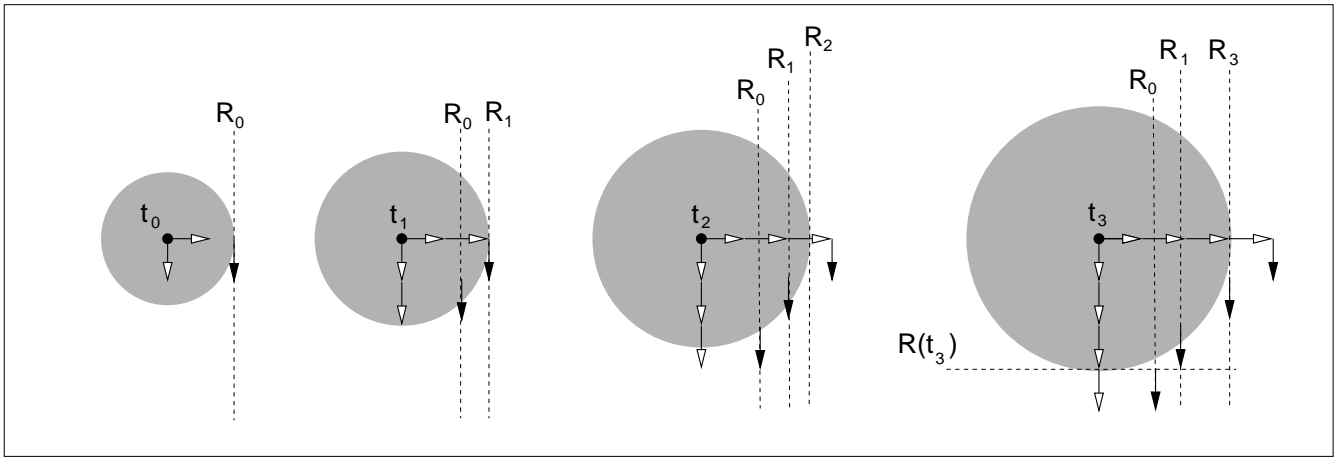}
\caption{\label{fig1}Schematic illustration of the geometry of Ly$\alpha$
and redshifted 21cm observations of a relativistically expanding quasar
Str\"omgren sphere (dark regions).  The four panels show the evolution at 4
different times. The quasar turns on at $t_0$, emitting photons
isotropically into a pre-existing \ion{H}{2} region of radius $R_0$. Ly$\alpha$ 
photons (open arrows) emitted at $t_0$ reach the edge of the \ion{H}{2} region at
$t_1$. At that time, 21cm photons (closed arrows) are emitted by the heated
gas perpendicular to the line-of-sight (at the quasar redshift) at 
$R>R_0$ and travel towards the observer. At $t_3$, the Ly$\alpha$
photons emitted at $t_1$ reach the edge of the \ion{H}{2} region. The absorption
of these photons encodes a measurement of $R_3$. When these photons are
observed, they are accompanied by 21 cm photons (emitted at
$t_1$ from the quasars redshift), which show a ring of radius $R_0$. }
\end{figure*}

The most natural place to search for the redshifted 21cm signature of
quasar \ion{H}{2} regions is around the highest redshift quasars discovered
in wide field surveys such as the {\it Sloan Digital Sky Survey}
(SDSS\footnote{See http://www.sdss.org/}). Such quasars have \ion{H}{2}
regions that are also probed via their Ly$\alpha$ absorption spectra. In
this section we discuss the complimentary nature of the two observations
with respect to the determination of the quasar lifetime, $t_{\rm age}$,
and the neutral fraction of the IGM, $x_{\rm HI}$.

Str\"omgren spheres expand faster into a partially ionized IGM. If the IGM
has a neutral fraction $x_{\rm HI}$, then the observation of $R_{\rm Ly}$
through Ly$\alpha$ absorption measures the combination $(t_{\rm
age,e}/x_{\rm HI})^{1/3}$ (where here we have neglected recombinations and
assumed $R_0=0$ for simplicity). Detection of the redshifted 21cm ring
reflects the state of the \ion{H}{2} region at $t_{\rm age,e}$. The
differential antenna temperature $\delta T_{\rm b}$ is proportional to
$x_{\rm HI}$ due to the small optical depth at 21 cm,
\begin{equation}
\delta T_{\rm b}\propto (1-e^{-\tau_{\rm HI}})\approx \tau_{\rm HI}\propto
x_{\rm HI}.
\end{equation}
Measurement of the contrast in the emission and absorption rings, or of the
brightness temperature of the emission ring, will facilitate a measurement
of $x_{\rm HI}$ independent from $t_{\rm age,e}$. Redshifted 21cm
observations may therefore provide an opportunity to break the degeneracy
between $t_{\rm age,e}$ and $x_{\rm HI}$. Unfortunately, the heating rate
and temperature of the IGM are uncertain, and unless $T_{\rm k}\gg T_{\rm
CMB}$, the value of $\delta T_{\rm b}$ will be sensitive to the unknown
value of $T_{\rm k}$. Below we discuss an alternative geometric method for
finding $x_{\rm HI}$ that is not sensitive to the precise value of $\delta
T_{\rm b}$.

The relativistic expansion and finite light travel time imply that the 21cm
and Ly$\alpha$ measurements sample two different epochs in the evolution of
the \ion{H}{2} region. The situation is illustrated schematically in
figure~\ref{fig1}. Suppose that the quasar turns on at $t_0$, emitting
photons isotropically into an existing \ion{H}{2} region of radius
$R_0$. Ly$\alpha$ photons emitted at $t_0$ will reach the edge of the \ion{H}{2} region at
$t_1$. At that time, 21cm photons with $\delta T_{\rm b}>0$ are emitted by
the heated gas at $R>R_0$ and travel towards the observer. 
For simplicity consider 21cm photons that are emitted from the
heated IGM just beyond the ionizing front and perpendicular to the 
line-of-sight (i.e. at the quasar redshift). The sphere
continues to expand at nearly the speed of light between $t_1$ and
$t_3$. At $t_3$, the Ly$\alpha$ photons emitted at $t_1$ reach the edge of
the \ion{H}{2} region. The absorption of these photons encodes a measurement of
$R_3$. The continuum and 21cm photons propagate to the observer, who
measures a radius $R_3$ along the line-of-sight in Ly$\alpha$ absorption,
and a different radius $R_0$ in redshifted 21cm photons. 

In this paper we define $R_{\rm Ly}$ to be the radius measured through
Ly$\alpha$ absorption along the line-of-sight. This is compared with
$R_{\rm HI}$, defined to be the radius of the ring measured perpendicular
to the line of sight through redshifted 21cm emission. The radii $R_{\rm
Ly}$ and $R_{\rm HI}$ are not equal for a spherical region due to finite
light travel time effects mentioned in the previous paragraph. We note that
this effect is separate from the super-luminal transverse expansion of the
ring of 21cm emission from the relativistically expanding portions of the
\ion{H}{2} region near the line-of sight.  These super-luminally expanding
rings would be seen $\sim3000$km$\,$s$^{-1}$ blueward of the quasar
redshift (for $R_{\rm p}\sim4$Mpc).  

\begin{figure*}[htbp]
\epsscale{1.5}
\plotone{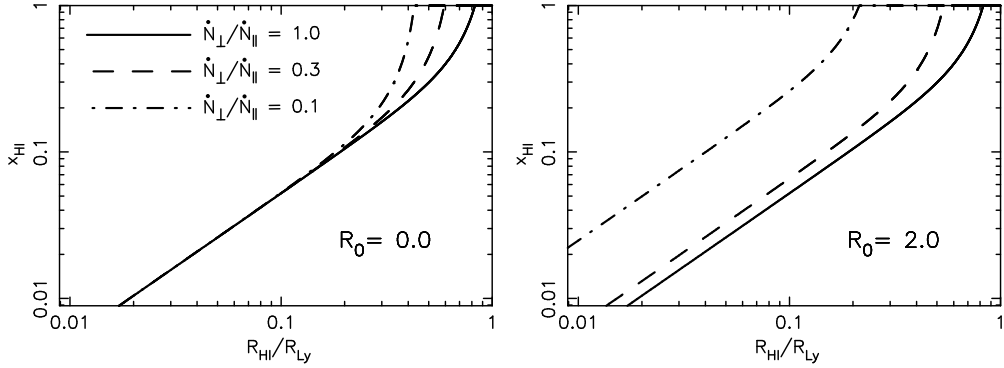}
\caption{\label{fig2} The neutral fraction as a function of the ratio
$R_{\rm HI}/R_{\rm Ly}$ calculated from equation~(\ref{neutfrac}). The
calculations are for the case where $R_{\rm Ly}=4.5$Mpc, $z=6.5$ and
$\dot{N}_\parallel =10^{57}$s$^{-1}$. We show curves for
$\dot{N}_{\perp}/\dot{N}_\parallel =1.0,0.3$ and 0.1, and for $R_0=0$
(left hand panel) and $R_0=2$ (right hand panel).}
\end{figure*}

The exact redshift of the quasar may be uncertain by up to
$\sim1000$km$\,$s$^{-1}$ (Richards et al.~2002).  However this uncertainty
corresponds to only 30\% of the diameter of the \ion{H}{2} region
($\sim3000$km$\,$s$^{-1}$). A point on the edge of the \ion{H}{2} region
that is blueshifted by $\sim1000$km$\,$s$^{-1}$ relative to the quasar lies
in a direction away from the quasar that makes an angle of
$\beta\sim\cos^{-1}(1000\mbox{km}\,
\mbox{s}^{-1}/3000\mbox{km}\,\mbox{s}^{-1})\sim70^\circ$
with the line-of-sight.  As a result, the misidentification of the quasar
redshift will lead to an error of $\sim R_{\rm p}(1-\sin\beta)\sim
0.05R_{\rm p}$ in the transverse diameter of the ring at the quasar
redshift.  Moreover, since $\beta\ga70^\circ$, the ring of 21cm emission
at the observed quasar redshift will not be subjected to superluminal
expansion, even if the quasar redshift is misidentified by
$\sim1000$km$\,$s$^{-1}$.

More quantitatively, the relativistic expansion and finite light travel time
lead to several interesting possibilities:

\begin{itemize}

\item The coupling of the spin temperature of neutral hydrogen to the
temperature of the warm IGM surrounding the \ion{H}{2} region results in
redshifted 21cm radiation in excess of the CMB.  $R_{\rm HI}$ is measured
transverse to the line-of-sight and so corresponds to the radius at the
time the observed UV photons were emitted. We can directly infer the
ionizing photon rate ($\dot{N}_\parallel $) which drives the evolution of
the \ion{H}{2} region along the line of sight. However, since the quasar
emission may not be isotropic, we allow the evolution of the \ion{H}{2}
region transverse to the line of sight to be driven by a different ionizing
photon rate $\dot{N}_{\perp}$. Thus, the ratio
$\dot{N}_{\perp}/\dot{N}_\parallel $ is a measure of the anisotropy in
quasar emission. Since high-redshift quasars are brightness selected, we
assume that $\dot{N}_{\perp}/\dot{N}_\parallel \leq1$.  Using
equation~(\ref{evinf}), we find the age of the quasar when light crossing
the edge of the \ion{H}{2} region was emitted ($t_{\rm age,e}$) and
substitute it into equation~(\ref{SE1}) for the radius of the \ion{H}{2}
region {\em at that time}. This yields the following relationship between
$R_{\rm Ly}$ and $R_{\rm HI}$
\begin{eqnarray}
\label{RR1}
\nonumber R_{\rm Ly}^{3} &=&
R_0^3\left(\frac{\dot{N}_\parallel }{\dot{N}_{\perp}}-1\right) + R_{\rm
HI}^3\left(\frac{\dot{N}_\parallel }{\dot{N}_{\perp}}\right)\\ &&\hspace{10mm} +
\frac{R_{\rm HI}}{c}\frac{3}{4\pi}\frac{\dot{N}_\parallel }{x_{\rm HI}n_{\rm
H}^0(1+z)^3},
\end{eqnarray}
or
\begin{eqnarray}
\label{RR}
\nonumber \left(\frac{R_{\rm HI}}{R_{\rm
Ly}}\right)^3&=&\left(\frac{\dot{N}_{\perp}}
{\dot{N}_\parallel }\right)-\left(\frac{R_0}{R_{\rm
Ly}}\right)^3\left(1-\frac{\dot{N}_{\perp}}{\dot{N}_\parallel }\right)\\ \nonumber
&-&\left[\left(\frac{R_{\rm HI}}{R_{\rm Ly}}\right)\left(\frac{R_{\rm
Ly}}{3.25\mbox{Mpc}}\right)^{-2}\right.\\ &&\left.\times x_{\rm
HI}^{-1}\left(\frac{\dot{N}_{\perp}}
{10^{57}\mbox{s}^{-1}}\right)\left(\frac{1+z}{7.5}\right)^{-3}\right].
\end{eqnarray}
Note that equations~(\ref{RR1}) and (\ref{RR}) are independent of $R_0$
when the quasar emission is isotropic, and that the dependence of $R_{\rm
HI}/R_{\rm Ly}$ on $\dot{N}_{\perp}/\dot{N}_\parallel $ is rather mild.
If the quasar emission is isotropic, the transverse radius $R_{\rm HI}$ is
always smaller than the line-of-sight radius $R_{\rm Ly}$, particularly
during the relativistic expansion phase.  When the expansion becomes
sub-relativistic, $R_{\rm HI}$ approaches $R_{\rm Ly}$.

\item The radii measured through Ly$\alpha$ absorption and 21cm emission 
correspond to different epochs separated by $R_{\rm Ly}/c$. For a 
spherical \ion{H}{2} region we obtain the average expansion
velocity of the ionizing front between the times corresponding to 
$R_{\rm HI}$ and $R_{\rm Ly}$ 
\begin{equation}
\langle\frac{dR_{\rm p}}{dt}\rangle=c\left(1-\frac{R_{\rm HI}}{R_{\rm
Ly}}\right).
\end{equation} 
Note that this velocity is the physical velocity of the ionizing front.
The apparent transverse velocity of portions of the ring might be
superluminal.

\begin{figure*}[htbp]
\epsscale{.8}
\plotone{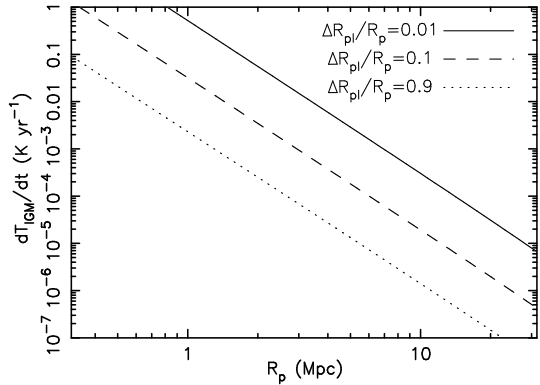}
\caption{\label{fig3}Plots of the heating rate of the IGM $dT_{\rm IGM}/dt$
(in K$~{\rm yr^{-1}}$) due to the X-ray emission by the quasar, as a
function of the Str\"omgren radius.  The three curves correspond to radii
at which the photons have traversed \ion{H}{1} path lengths of $\Delta
R_{\rm pl}=0.01R_{\rm p}$, 0.1$R_{\rm p}$ and 0.9$R_{\rm p}$, where $R_{\rm
p}$ is the physical radius of the \ion{H}{2} region.  The assumed spectrum
is given in equation~(\ref{spec}), and the neutral fraction $x_{\rm HI}$
was set to unity.}
\end{figure*}

\item Given a fixed luminosity, the expansion speed at a radius $R_{\rm
Ly}$ is proportional to $x_{\rm HI}^{-1/3}$, and so the neutral fraction
can be inferred from the two radii.  
Equation~(\ref{RR}) yields
\begin{eqnarray}
\nonumber
\label{neutfrac}
x_{\rm HI}&=&\left(\frac{R_{\rm Ly}}{c}\frac{\dot{N}_\parallel }{n_{\rm
H}^0(1+z)^3}\right)\\ \nonumber
&&\hspace{-10mm}\times\left[\frac{4\pi}{3}\left(R_{\rm Ly}^3-R_{\rm
HI}^3\left(\frac{\dot{N}_\parallel }{\dot{N}_{\perp}}
\right)-R_0^3\left(\frac{\dot{N}_\parallel }{\dot{N}_{\perp}}
-1\right)\right)\right]^{-1}\\ \nonumber &=&\left[\frac{\left({R_{\rm
HI}}/{R_{\rm Ly}}\right)}{1-\left(\frac{R_{\rm HI}}{R_{\rm
Ly}}\right)^3\left(\frac{\dot{N}_\parallel }{\dot{N}_{\perp}}\right) -
\left(\frac{R_0}{R_{\rm
Ly}}\right)^3\left(\frac{\dot{N}_\parallel }{\dot{N}_{\perp}}
-1\right)}\right]\\ &&\hspace{-0mm}\times\left(\frac{R_{\rm
Ly}}{3.25\mbox{Mpc}}\right)^{-2}
\left(\frac{\dot{N}_\parallel }{10^{57}\mbox{s}^{-1}}\right)
\left(\frac{1+z}{7.5}\right)^{-3} .
\end{eqnarray}
Some examples of the dependence of $x_{\rm HI}$ on the ratio $(R_{\rm
HI}/R_{\rm Ly})$ given $R_{\rm Ly}=4.5$Mpc and
$\dot{N}_\parallel =10^{57}$s$^{-1}$, are shown in figure~\ref{fig2}. In
the case of isotropic quasar emission
($\dot{N}_{\perp}/\dot{N}_\parallel =1$), a small ratio of $(R_{\rm
HI}/R_{\rm Ly})$ implies relativistic expansion, and hence a small neutral
fraction unless $R_{\rm Ly}\ll3.25$Mpc. Values of
$\dot{N}_{\perp}/\dot{N}_\parallel$ which are less than unity lead to
larger inferred values of $x_{\rm HI}$. Similarly, a non-zero value for
$R_0$ leads to larger values of $x_{\rm HI}$. The assumptions of
isotropy and $R_0=0$ lead to determination of a lower limit for $x_{\rm
HI}$. The observation of a redshifted 21cm ring with a radius greater than
2.5Mpc (0.5Mpc) around the known $z\ga 6.3$ quasars (with $R_{\rm
Ly}\sim4.5$Mpc) would provide a limit on the neutral fraction of $x_{\rm
HI}>0.3$ ($x_{\rm HI}>0.05$).

\end{itemize}

\section{Some Examples}
\label{ex}

In this section we present examples of the redshifted 21cm signatures of
$z>6$ quasars that will be observed by instruments like {\em LOFAR}.  The
UV continuum blueward of Ly$\alpha$ is assumed to be that of the median
spectrum in the sample of HST quasars studied by Telfer, Zheng, Kriss \&
Davidsen~(2002).  They found a power-law $\epsilon_{\rm c}(\nu)\propto
\nu^{-\alpha_{\rm EUV}}$ with $\alpha_{\rm EUV}=1.57$. This slope may be
extended into the X-ray regime based on the results of Yuan, Brinkmann,
Siebert \& Voges~(1998). We include the Ly$\alpha$ emission line centered
at $\lambda_\alpha=912$\AA, with an equivalent width of $W_\lambda=90$\AA
(Vanden Berk et al.~2001; Telfer, Zheng, Kriss \& Davidsen~2002), and a
line width of $\sigma_\lambda=19$\AA (Vanden Berk et al.~2001), providing a
UV--X-ray spectrum
\begin{eqnarray}
\nonumber
\label{spec}
\epsilon(\nu) &=& 1.3\times10^{31}\mbox{erg}\,\mbox{s}^{-1}\,\mbox{Hz}^{-1}\left(\frac{\nu}{\nu_{\rm ion}}\right)^{-\alpha_{\rm EUV}}\\
&\times&\left[1+\frac{1}{\sqrt{2\pi}\sigma_\lambda}\exp{\left(\frac{(c/\nu-\lambda_\alpha)^2}{\sigma_\lambda^2}\right)}\right].
\end{eqnarray}
The normalization of this spectrum is appropriate for a
$2\times10^9$M$_\odot$ supermassive black hole accreting at its Eddington
rate (Elvis et al.~1994), and it yields an ionizing photon rate of
$\dot{N}_\parallel =1.3\times10^{57}$s$^{-1}$, representative of the $z\ga
6.3$ SDSS quasars (White et al.~2003). In this section, we show examples of
the 21cm emission signal that may be observed around the known very high
redshift quasars, with the spectrum described by equation~(\ref{spec}).

The level of redshifted 21cm emission is enhanced through heating of the
IGM by the quasar X-ray emission.  The corresponding heating rates $dT_{\rm
IGM}/dt$ of the above quasar spectrum are plotted as a function of radius
in figure~\ref{fig3}.  The three curves correspond to radii where photons
traversed path lengths through neutral gas of $\Delta R_{\rm pl}=0.01R_{\rm
p}$, 0.1$R_{\rm p}$ and 0.9$R_{\rm p}$.  The neutral fraction was set to
unity in these examples. Note that lower neutral fractions lead to higher
values of $f_{\rm x}$ and hence larger heating rates. 

\begin{figure*}[htbp]
\epsscale{1.5}
\plotone{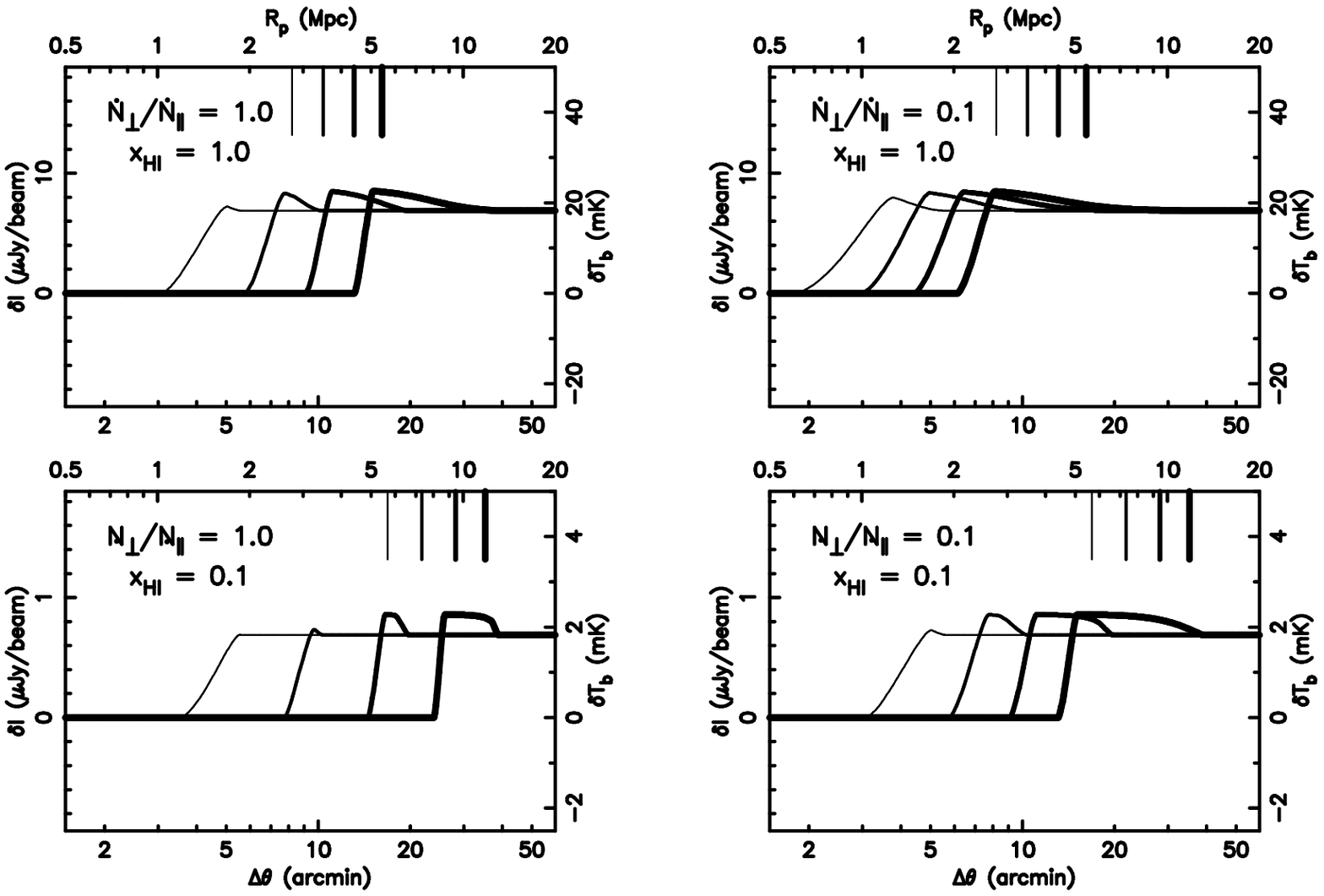}
\caption{\label{fig4} Four examples of the evolution of a Str\"omgren
sphere expanding into a warm ($T_{\rm IGM}=5T_{\rm CMB}$) IGM as seen in
redshifted 21cm emission and Ly$\alpha$ absorption, assuming different
combinations of $\dot{N}_{\perp}/\dot{N}_\parallel =1.0,0.1$ and $x_{\rm
HI}=1.0,0.1$.  In all cases we assume $R_0=0$.  The 21cm emission signal is
plotted as a function of radius (upper horizontal axes) and apparent angle
in arc-minutes (lower axes). 
The y-axes are labeled by the differential antenna
temperature $\delta T_{\rm b}$ (right axes)
and the flux $\delta I$ (left axes) given a circular top-hat beam with a
diameter of $2^\prime$. In each case, four profiles are shown for observing
times corresponding to quasar ages of $t_{\rm age,e}=0.5,1,2,4\times10^7$yr
when the {\em optical photons} were emitted.  Values of $R_{\rm Ly}$ at
$t_{\rm age,e}=0.5,1,2,4\times10^7$yr are shown with tick marks.}

\vspace{15mm}
\epsscale{1.5}
\plotone{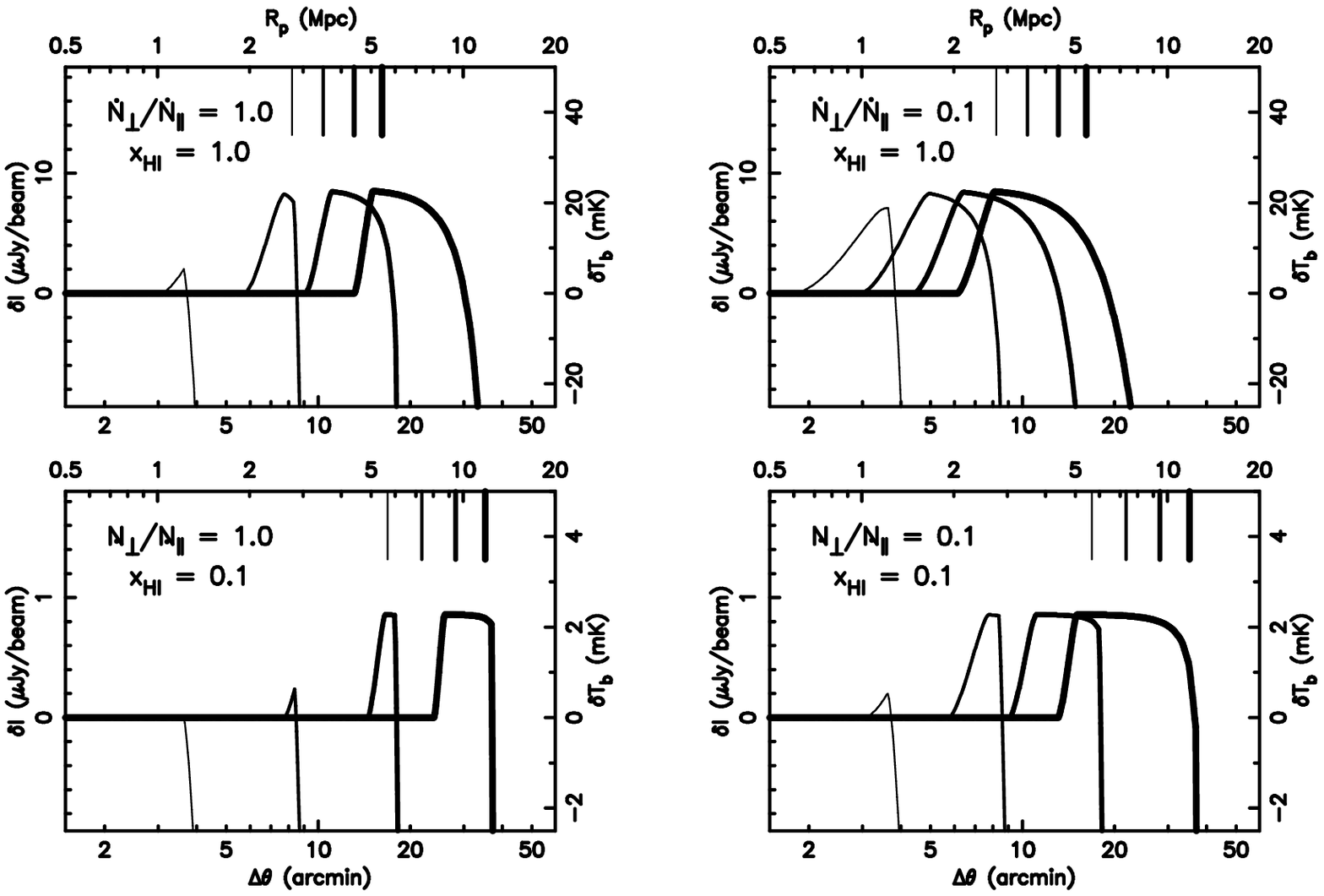}
\caption{\label{fig5} Same as in figure~\ref{fig4}, but with a cold IGM
having $T_{\rm IGM}=0.026(1+z)^2\ll T_{\rm CMB}$.}
\end{figure*}

\begin{figure*}[htbp]
\epsscale{.8}
\plotone{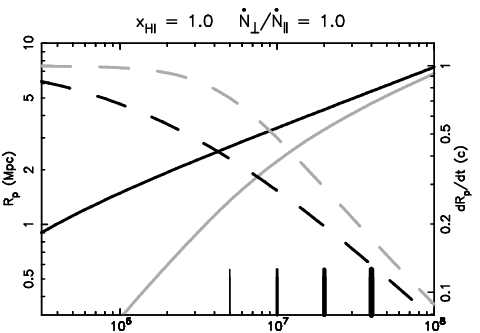}
\caption{\label{fig6} Plots of the radius measured through Ly$\alpha$
absorption along the line-of-sight $R_{\rm Ly}(t_{\rm age})$ (solid dark
lines) as well as the radius of the interior boundary of the 21cm emission
ring at the quasar redshift (solid light lines).  The dashed lines show the
speed at which the edge of the \ion{H}{2} region seen in 21cm (light lines)
and Ly$\alpha$ absorption (dark lines) is moving when observed at a time
corresponding to $t_{\rm age,e}$.  The curves correspond to the example
shown in the top left panel of figures~\ref{fig4} with
$\dot{N}_{\perp}/\dot{N}_\parallel =1$ and $x_{\rm HI}=1$.}
\end{figure*}

If the Str\"omgren spheres expand into a pre-heated IGM containing a UV
background that coupled the spin temperature of the IGM to its kinetic
temperature, the \ion{H}{2} regions should appear as holes in the sky of
redshifted 21cm emission.  As an example we first consider an IGM that is
pre-heated to $\sim100$K at $z=6.3$ 
($T_{\rm IGM}\sim5T_{\rm CMB}$).  We follow the evolution of the
Str\"omgren sphere, and at each time compute the heating rate of the IGM
beyond the edge of the \ion{H}{2} region using equation~(\ref{heat}).  The
evolution of the kinetic temperature of the IGM is followed on a grid using
equation~(\ref{temp}). The spin temperature is then set to $T_{\rm IGM}$ as
appropriate in the presence of a strong Ly$\alpha$ background.
Figure~\ref{fig4} shows four examples of the 21 cm emission signature of a
Stromgren sphere expanding into a warm IGM as a function of radius (top
axes) and apparent angle (bottom axes) for different combinations of
$\dot{N}_{\perp}/\dot{N}_\parallel =1.0,0.1$ and $x_{\rm HI}=1.0,0.1$. An
initial value of $R_0=0$ was assumed.  The y-axes are labeled by the
differential antenna temperature $\delta T_{\rm b}$ (right axes), and the
flux $\delta I$ (left axes) assuming the design goal sensitivity 
of {\it LOFAR}
\begin{equation}
\delta I = (1.2\mu\mbox{Jy}\,\mbox{arcmin}^{-2})\left(\frac{\delta T_{\rm
b}}{10\mbox{mK}}\right)\left(\frac{1+z}{7.5}\right)^{-2},
\end{equation}
and a top-hat beam with a diameter of $2^{\prime}$. In each case, four
profiles are shown for observing times corresponding to quasar ages of
$t_{\rm age,e}=0.5,1,2,4\times10^7$yr at the time when the {\em optical
photons} observed along the line-of-sight were emitted. The lower row
includes cases with $x_{\rm HI}=0.1$.  The smaller neutral fraction results
in larger radii, faster relativistic expansion, and therefore thinner
emission shells.  The 21cm emission signal is weaker in this case because
$\delta T_{\rm b}\propto x_{\rm HI}$. There is excess emission above the
level of the general IGM in a ring covering the region where gas has been
heated by the quasar. However the brightness temperature is only slightly
higher than the rest of the IGM. This is due to the temperature dependence
of $\Delta T_{\rm b}$ which asymptotes to a constant value of $\sim23x_{\rm
HI}$mK for high temperatures. The panels on the right show examples with
$\dot{N}_{\perp}/\dot{N}_\parallel =0.1$. The lower transverse flux
results in an expansion that becomes sub-relativistic at earlier times.  As
a result, the emission shell has a smaller radius and a larger thickness.

As our second example we consider an IGM that is cold in neutral regions
[$T\sim0.026(1+z)^2$, corresponding to adiabatic cooling following the
thermal decoupling of the baryons from the CMB]. The resulting 21 cm signal
is shown in figure~\ref{fig5}. The heated portions of the IGM beyond the
edge of the \ion{H}{2} region show emission, while the cold IGM produces
absorption of the CMB at larger radii. The brightness temperature of the
emission ring obtains values of $T_{\rm b}\sim20x_{\rm HI}$mK, smaller in
amplitude than the absorption signal of $T_{\rm b}\sim-295x_{\rm HI}$mK. A
cold IGM therefore offers the best contrast for observation of the
transverse extent of the \ion{H}{2} region.  As in the previous example,
the emission rings get thicker at late times because the expansion speed
becomes less relativistic. Since the transverse expansion becomes
sub-relativistic sooner than the line-of-sight expansion, the rings become
thick at earlier times for a lower value of
$\dot{N}_{\perp}/\dot{N}_\parallel $.

Figure~\ref{fig6} shows $R_{\rm Ly}(t_{\rm age})$ (solid dark lines) as
well as the inner radius of the 21cm emission ring (solid light lines) for
the example presented in the top left panels of figures~\ref{fig4} and
\ref{fig5} ($\dot{N}_{\perp}/\dot{N}_\parallel =1$, $x_{\rm HI}=1$).  The
four epochs corresponding to the differential antenna temperature profiles
plotted in figures~\ref{fig4} and \ref{fig5} are indicated by the vertical
tick marks.  Similarly, values of $R_{\rm Ly}$ at $t_{\rm
age,e}=0.5,1,2,4\times10^7$yr are indicated by tick marks in the panels of
figures~\ref{fig4} and \ref{fig5}. The value of $R_{\rm Ly}$ is always
larger than the radius of 21cm emission. The dashed lines show the velocity
at which the edge of the \ion{H}{2} region is moving when observed at a
time corresponding to $t_{\rm age,e}$. The 21cm ring (light dashed lines)
is observed at an earlier phase of evolution and so is moving at a higher
velocity. The edge of the \ion{H}{2} region is moving at a speed
$\ga 0.2c$ when $R_{\rm p}\la 4.5$Mpc.

\section{Fossil \ion{H}{2} regions in the high redshift IGM}
\label{fossil}

The presence of a Ly$\alpha$ background has very interesting implications
for the observation of fossil \ion{H}{2} regions. The recombination time
within an ionized region of the IGM at mean density is $t_{\rm recom}\sim
(\alpha_{\rm B} n_{\rm H}^0(1+z)^3)^{-1}\sim1.5\times10^9[(1+z)/7.5]^{-3}$
years, where $\alpha_{\rm B}=2.6\times10^{-13}\,$cm$^3\,$s$^{-1}$ is the
case-B recombination coefficient at a temperature of $10^4$K.  Thus, the
recombination time at the mean cosmic density exceeds the Hubble time for
$z\la 7$. 
Once a quasar has turned off, it therefore leaves behind a fossil
\ion{H}{2} region that survives recombination. This \ion{H}{2} region will
therefore remain detectable in redshifted 21cm radiation in both scenarios
discussed in this paper.  Since the bright SDSS quasars are likely to be
active for only $\sim 1\%$ of the time (Wyithe \& Loeb~2003b), there should
be of order 100 times as many 21cm emission rings around fossil \ion{H}{2}
regions as around active quasars.  Furthermore, the hierarchical buildup of
the supermassive black holes powering the quasars would result in an even
larger number of small fossil \ion{H}{2} regions.

How many \ion{H}{2} regions should an instrument like {\it LOFAR} detect?  It is
anticipated that near frequencies of $\nu_{\rm bp}\sim200$MHz, the virtual
core of the planned instrument {\em LOFAR} will have a beam-size of
$\Delta\theta_{\rm beam}\sim23^\circ$, a resolution of several tens of
arc-minutes, and a system temperature of $T_{\rm sys}\sim300$K. The
instrument will have an effective collecting area of $A\sim10^5$m$^2$ at
this frequency, leading to a resolution of $\Delta\theta_{\rm
fa}\sim1.22\lambda\sqrt{\pi/A}\sim22'$ for the telescope in a filled aperture
mode.  Higher resolutions are obtained by spreading the array out over
longer baselines, at the cost of decreased sensitivity. From the radiometer
equation (e.g. Burke \& Graham-Smith~1997) we can find the root-mean-square
noise in the brightness temperature as a function of resolution on
angular scales $\Delta\theta<\Delta\theta_{\rm fa}$ of
\begin{eqnarray}
\nonumber\Delta T_{\rm b}(\Delta\theta) &=&(0.24
\mbox{mK})\left(\frac{\Delta\theta}{\Delta\theta_{\rm
res}}\right)^2\left(\frac{T_{\rm sys}}{300\mbox{K}}\right)\\
&&\hspace{2mm}\times\left(\frac{\Delta\nu_{\rm
bp}}{4\mbox{MHz}}\right)^{-1}\left(\frac{\tau}{100\mbox{hr}}\right)^{-1/2}\\
\end{eqnarray}
Here $\Delta \nu_{\rm bp}$ is the detector band-pass, and $\tau$ is the
integration time. The noise in the brightness temperature is constant on
scales larger than $\Delta \theta_{\rm fa}$. The root-mean-square noise for
an observation with a resolution $\Delta\theta\sim3'$ therefore becomes
comparable to the expected emission signal ($\delta T_{\rm b}\sim10$mK) for
a modest integration time of $\tau\sim100$hr.

The co-moving optical luminosity function of bright quasars at $z\sim6$ in
units of $L_\odot^{-1}$Gpc$^{-3}$ (Fan et al.~2003) is
\begin{equation}
\Phi(L)\approx
10^{-13}L_\odot^{-1}\mbox{Gpc}^{-3}
\left(\frac{L}{10^{13.1}L_\odot}\right)^\beta,
\end{equation}
where $\beta\sim-3$ is the logarithmic slope at the bright end. Given a
band-pass $\Delta \nu_{\rm bp}\sim4$MHz and a central frequency $\nu_{\rm
bp}\sim200$MHz, the number of active quasars per logarithm of luminosity
per field of view is
\begin{equation}
\label{N}
N\sim L\Phi(L)\frac{d^2V}{dzd\Omega}\pi\left(\frac{\Delta\theta_{\rm
beam}}{2}\right)^2(1+z)\frac{\Delta\nu_{\rm bp}}{\nu_{\rm bp}},
\end{equation}
where ${d^2V}/{dz~d\Omega}$ is the co-moving volume per unit redshift per
unit solid angle.  Equation~(\ref{N}) yields a value of $N\sim2$ bright
$z\sim6$ quasars per {\it LOFAR} field.

The number of redshifted 21cm rings or holes in the emission signature of
the IGM per field as determined by equation~(\ref{N}) applies to active
quasars. However as noted above, the 21cm emission may persist after the
quasar has turned off. This is because in the absence of other heating
sources, gas heated to temperatures far in excess of the CMB will cool
adiabatically. The $(1+z)^2$ dependence of the temperature in this regime
should ensure that the gas remains at a temperature in excess of the CMB by
the time it is re-heated during the next episode of quasar activity. We
might expect to find significant numbers of fossil \ion{H}{2} regions,
provided that there is continued coupling of $T_{\rm s}$ and $T_{\rm IGM}$
due to a UV background. The number of detectable fossil \ion{H}{2} regions
would then equal the number in equation~(\ref{N}) multiplied by the inverse
of the quasar duty cycle $\sim1\%$ (Wyithe \& Loeb~2003b), or about
200 per field.

{\em LOFAR} should also be sensitive to \ion{H}{2} regions around less
luminous quasars. The brightness temperature of the emission ring, or of
the IGM outside the \ion{H}{2} region has a mild dependence on the quasar
luminosity, and its diameter scales only with the one third power of the
luminosity. \ion{H}{2} regions around less luminous quasars should be more
numerous by a factor proportional to $L^{\beta+1}$. For example, detection
of redshifted 21cm emission around quasars with luminosities a factor of 10
smaller than those of the bright SDSS quasars require a resolution that is
better by a factor of $10^{1/3}$, corresponding to an integration time that
is longer by a factor of $10^{2/3}\sim4.6$. However, the number of active
high redshift quasars per field powering such emission rings is likely to
be of order $10^{2}N\sim200$, while the number of fossils regions could be
$\sim2\times10^4$.

Finally, we briefly discuss the example of a redshifted 21cm signature
around a high redshift quasar presented in 5b of Tozzi et al.~(2000).
Their example shows a ring of redshifted 21cm emission inside a ring of
redshifted 21cm absorption. This morphology is unlikely to be observed
because the observation of an absorption ring requires the absence of a UV
background, and the provision of coupling of the spin and kinetic
temperatures through Ly$\alpha$ photons from the quasar itself. As we have
shown, Ly$\alpha$ photons from the quasar are unable to couple the spin and
kinetic temperatures.  An emission ring enclosed within an IGM that absorbs
everywhere, as shown in figure~\ref{fig5} is the correct signature of a
Str\"omgren sphere expanding into a cold IGM (provided that there is a UV
background).

\section{Str\"omgren spheres and the global signature of reionization}
\label{global}

A very exciting possibility for instruments like {\em LOFAR} is to search
for the emission (or absorption) of redshifted 21cm radiation from the
neutral IGM (averaged over large scales) before reionization. 
By finding the frequency at which this
emission (absorption) cuts off, one might determine the redshift of
reionization (e.g. Shaver et al.~1999). If reionization took place over an
extended period of time as suggested by recent work (Barkana \& Loeb 2003b),
then this global reionization signature will be difficult to detect for two
primary reasons: (i) because it will be superimposed on a smooth but varying
foreground spectrum, and (ii) calibration needs to be performed over
a frequency range that is broader than the detector band-pass.

Arc-minute resolution observations of quasar Str\"omgren spheres would 
provide an alternative method to search for a step in the brightness 
temperature of the IGM due to reionization. This is because
the inner part of the \ion{H}{2} region provides a calibration of the
brightness temperature of the radio foreground that is unavailable along
other lines of sight.  Note that 21cm emission from the quasar itself does
not affect this calibration, since it effectively just adds to the
foreground.  Observations along the line-of-sight to the quasar would
include emission within the spectral ranges corresponding to both the
ionized region and the neutral gas within the same band-pass.  Redshifted
21cm observations of quasars at a series of redshifts would therefore allow
determination of the brightness temperature of the IGM at different cosmic
times, and a detection of the onset of reionization.
Redshifted 21cm observation of highest-redshift SDSS quasars will
immediately determine whether the IGM was significantly neutral at $z>6.3$.

\section{Discussion}
\label{disc}

Observations of the reionization epoch in redshifted 21cm radiation with
proposed instruments like {\it LOFAR} will offer a unique opportunity to
study the state of the IGM at the end of the dark ages.  The warm neutral
hydrogen outside the \ion{H}{2} regions surrounding bright high redshift
quasars is expected to emit at 21cm. The brightness temperature contrast of
this emission will probe the neutral fraction (with a lower contrast
implying a lower neutral fraction), although the inference will be subject
to uncertainties in the temperature of the warm IGM. Alternatively, the
relativistic expansion of the \ion{H}{2} regions provides a geometric
method to determine the neutral fraction as well as the quasar lifetime.

We have combined the observed Ly$\alpha$ absorption signal with
expectations for redshifted 21cm signatures of the \ion{H}{2} regions
around the highest redshift SDSS quasars. Because the \ion{H}{2} regions
around the known $z\ga 6.3$ quasars are at a stage of relativistic
expansion, the line-of-sight and transverse radii (measured through
Ly$\alpha$ absorption and redshifted 21cm emission, respectively),
correspond to two different epochs in the evolution of the \ion{H}{2}
regions. We have found that measurements of these two radii can be used to
determine the neutral fraction of the IGM with only a mild dependence on
the anisotropy of the quasar emission (see Fig. 2).  Assuming isotropic
quasar emission and the absence of a fossil \ion{H}{2} region prior to the
quasar activity, the observation of a redshifted 21cm emission ring with a
radius greater than 2.5Mpc (0.5Mpc) around the known $z\ga 6.3$ quasars,
would provide a lower limit on the neutral fraction of $x_{\rm HI}>0.3$
($x_{\rm HI}>0.05$).  Emission concentrated close to the line-of-sight and
the existence of fossil \ion{H}{2} regions around the quasar host galaxy
would imply even larger neutral fractions.

At low redshifts, quasar lifetimes are bracketed by a variety of
techniques to the range $\sim10^6$--$10^8$yr (e.g. Yu \& Tremaine~2002;
Martini \& Weinberg~2001; Haiman \& Hui~2001; Jakobsen, Jansen, Wagner \&
Reimers~2003), with preferred values near $10^7$ years
(e.g. Martini~2003). The lifetime required for the \ion{H}{2} region to
expand to its observed size is proportional to $x_{\rm HI}^{-1/3}$. The
geometric determination of the neutral fraction can therefore be directly
used to determine also the quasar lifetime measured from the combination of
$R_{\rm p}$ and $x_{\rm HI}$. Thus, Ly$\alpha$ and redshifted 21cm
measurements of the expanding \ion{H}{2} regions around the highest
redshift quasars will allow a direct determination of quasar ages on a case
by case basis and constrain quasar formation models (e.g. Haiman \&
Loeb~2001).

Based on current number counts of bright high redshift quasars and the
preliminary design of {\it LOFAR}, we forecast that each {\it LOFAR} field
(of radius $\sim11^\circ$) will contain 
$\sim 2$ active quasars as bright as the observed SDSS
quasars within one band-pass of frequency ($\sim4$MHz). An integration time of
$\sim100$ hours should be sufficient to detect the rings of redshifted 21cm
emission from the warm IGM outside the quasar \ion{H}{2} regions if $x_{\rm
HI}=1$.  In the presence of a strong cosmic Ly$\alpha$ background, these
rings should remain visible after the quasar has turned off. This would
result in the detection of a large number of fossil \ion{H}{2} regions that
exceeds the number of active quasars by the inverse of the quasar duty cycle
($\sim 100$).

Arc-minute observations of quasar Str\"omgren spheres will provide an alternative
method for measuring the spectral step that makes the so-called global signature of
reionization (Shaver et al. 1999). The brightness temperature measured in
the spectral range corresponding to the ionized gas within the \ion{H}{2}
region provides a calibration of the foreground emission at the same
redshift.  This calibration will help overcome the difficulties associated
with measuring the global signature of an extended reionization epoch (as
suggested by the large electron scattering optical depth measured by {\em
WMAP}; Kogut et al. 2003) on top of a frequency dependent
foreground. Detection of the signal for a statistical sample of quasars
would help calibrate the expected scatter in the reionization redshift of
different regions in the universe (Barkana \& Loeb 2003b).

Finally, we emphasize that the redshifted 21cm observation of existing SDSS
quasars will determine whether the IGM was significantly neutral at $z>6.3$
as suggested by their observed Ly$\alpha$ spectra (Wyithe \& Loeb~2004).
When combined with the high value of the optical depth for electron
scattering measured by {\it WMAP} (Kogut et al.~2003), a significant
neutral fraction at $z\sim6.3$ would imply a complex reionization history
and possibly an early reionization phase driven by Population-III stars.

\acknowledgements 

This work was supported in part by NASA grant NAG 5-13292, and by NSF
grants AST-0071019, AST-0204514 (for A.L.).

\end{document}